\documentclass[
twocolumn,
prd,amsfonts,showpacs
]{revtex4}

\usepackage{bm,graphics,graphicx,epsfig,color}
\begin{document}
\title{Probing the non-linear structure of general relativity with 
black hole binaries}
\author{K.G. Arun$^1$, B.R. Iyer$^1$, M.S.S. Qusailah$^{1,2}$ and
B.S. Sathyaprakash$^3$} 
\affiliation{$^1$Raman Research Institute, Sadashivanagar, Bangalore
560\,080, India}
\affiliation{$^2$On leave of absence from Sana'a University, Yemen}
\affiliation{$^3$School of Physics and Astronomy, Cardiff University, 
Cardiff, CF24 3YB, UK}
\begin{abstract}
Observations of the inspiral of massive binary black holes (BBH) in the Laser 
Interferometer Space Antenna (LISA) and stellar mass binary black holes
in the European Gravitational-Wave Observatory (EGO) offer an unique 
opportunity to test the non-linear structure of general relativity. 
For a binary composed of two non-spinning black holes, the non-linear 
general relativistic effects depend only on the masses of the constituents.
In a recent letter, we explored the possibility of a test to determine
 all the post-Newtonian coefficients in the gravitational wave-phasing.
 However,  mutual covariances dilute the effectiveness of such a
 test. In this paper, we propose a more powerful test in which 
the various post-Newtonian coefficients in 
the gravitational wave phasing are systematically measured by treating 
three  of them  as independent parameters and demanding their 
mutual consistency.  LISA  (EGO) will observe BBH inspirals with
a signal-to-noise ratio of more than 1000 (100) and thereby 
test the self-consistency of each of the nine post-Newtonian coefficients that 
have so-far been computed, by measuring the lower order coefficients to
a relative accuracy of $\sim 10^{-5}$ (respectively, $\sim 10^{-4}$)
and the higher order coefficients to a relative accuracy in the range
$10^{-4}$-$0.1$ (respectively, $10^{-3}$-$1$).
\end{abstract}
\maketitle
Binary pulsar observations provide one of the most stringent methods to
test the strong field regime of gravity in general
relativity (GR) and its alternatives~\cite{DamTay92}. The test is possible since the
orbital dynamics of the binary is relativistic enough to allow the
measurement of effects due to gravitational radiation damping
 at the post-Newtonian order $(v/c)^5.$
Binary pulsar measurements are performed
by fitting the pulse arrival times to a relativistic `timing'
model~\cite{timing,DamTay92}, which is 
a function of the Keplerian parameters (orbital period, eccentricity
and the projected semi-major axis of the pulsar orbit)
and post-Keplerian (PK) parameters (the periastron advance, time-dilation and 
secular change of the orbital period). 
Two more PK
parameters, related to the  Shapiro-delay caused by the gravitational
field of the companion, can be measured if the orbit is seen
nearly edge-on. Different theories of
gravity  have different predictions for the values of the PK parameters
 as functions of the individual masses of the binary
constituents $m_1$ and $m_2$.
Thus,  a measurement of three or more PK parameters facilitates
a test by requiring consistency, within the observational
errors, in the estimation of the masses of the two bodies as
determined by the various parameters.
The most rigorous test possible so far is with the  most relativistic
binary pulsar PSR
J0737-3039 \cite{DPulsar}. Observed almost edge-on, it permitted the
measurement of five PK parameters, which together with an additional constraint
from the measurement of mass-ratio, determine and check the consistency of
the masses of the two pulsars in the $m_1$-$m_2$ plane~\cite{DPulsar}.

Although  radio binary pulsars are capable of testing certain lower
post-Newtonian (PN) order
general relativistic effects, such as the advance of the periastron
and the quadrupole approximation to the generation of gravitational
waves, they will, unfortunately, not be able to probe the strong field 
non-linear effects, such as the tails of gravitational waves~\cite{BS93}. 
This is because the PN expansion parameter is of order
$v \sim 10^{-3}$ -- far too small for the effects that first 
appear at higher post-Newtonian orders to play a significant role 
in radio observations of binary neutron stars.
Space- and ground-based gravitational wave detectors, such as the 
Laser Interferometer Space Antenna (LISA), Laser Interferometer 
Gravitational-Wave Observatory (LIGO), VIRGO and European 
Gravitational-Wave Observatory (EGO), will observe compact binary 
neutron stars and binary black holes (BBH) in the last stages of their
non-linear evolution, during which the parameter $v$ is two orders of magnitude
larger ($v \sim 0.2$-$0.4$) than it is for current radio observations
of such systems. 
For some of the rare (about once per year) inspiral events observed 
by LISA (EGO) the amplitude signal-to-noise ratio  could be as
large as 
3,000 (100). Such high SNR events will allow us to measure the parameters
of the signal and the source quite accurately, thereby allowing tests 
that were not feasible earlier.
Different tests of GR have been proposed by various authors using 
GW observations 
of the inspiralling compact binaries
~\cite{BSat,Will,BBW04} and contrasted with the binary pulsar
observations~\cite{DE98}.
These tests would necessitate an accurate parameter extraction~\cite{AISS04} 
scheme using the highest PN order waveform available.

The GW `phasing formula' is very close in spirit to the 
`timing formula' used in the binary pulsar observations.
The timing formula, $\phi_n^{\rm PSR}$=$F_{\rm T}[t_n,p_i]$, 
connects the rotational phase $\phi_n$ of a spinning pulsar
to the time-of-arrival $t_n$ of the radio signal and a set of Keplerian 
and PK parameters $p_i$=$\{p^{\rm K},p^{PK}\}$. Similarly,
a precise model for GWs from a compact binary will need  accurate
information about the continuous evolution of the GW phase. Schematically,
the phasing formula reads $\phi^{\rm GW}$=$F_{\rm P}[t,q_i]$ where, in
Einstein's theory, $q_i$
carry the information of the source via functions of the individual
masses and spins.
The phasing formula consists of different PN parameters $q_i$, 
similar to the PK parameters of the timing formula, and is currently
available up to relative 3.5PN order
{\it i.e.,} ${\cal O}(v^7)$~\cite{2PNphasing,phasing}.
In the present paper we propose 
and explore an interesting possibility of testing General Relativity with 
the high-SNR GW observations of BBH inspirals by LISA and EGO. The 
proposed test is similar in essence to the binary pulsar test, but in a
stronger and dynamic regime of gravity. 
Using the two lowest order PN coefficients $q_i$ as basic variables
to parametrize the waveform and choosing the other 
PN coefficients as `test' parameters, one at a time, it is possible to
perform  many consistency checks of the PN coefficients in the $m_1$-$m_2$ plane. 
In the rest of the paper we investigate this possibility in greater
detail.

Binary black holes in close orbit around each other are highly relativistic
and mandate the inclusion of higher PN  order terms in their description.
Gravitational waves emitted during the inspiral phase 
comprise a variety of terms arising from
the non-linear multipole interactions as the radiation propagates from
the source to the far-zone~\cite{tail}. 
These non-linear interactions lead to the phenomenon of tails at orders 
1.5PN and 2.5PN (propagation not only on but inside the
light cone as well)  and tails-of-tails at 3PN. 
For spinning binaries, there also exist effects of
 spin-orbit and spin-spin couplings at 1.5PN and 2PN, respectively.
These effects are imprinted in the emitted gravitational radiation 
and can be extracted by matching the detector data with an expected 
gravitational waveform, often called an optimal filter or a template. 
The template itself  can  only be
computed using post-Newtonian theory in which the various physical
quantities relevant to the emission of gravitational waves are
expanded in an asymptotic series in the small parameter $v$ -- the
characteristic velocity in the system~\cite{footGc}. An important feature  of the
PN expansion is the presence of $\log$-terms  $v^{m}(\ln v)^{n},$
where $m$ and $n$ are integers. General relativity is incompatible
with a simple Taylor expansion
in only powers of  $v.$ For instance,
currently, the expansions of the specific binding energy $E$ and
gravitational wave flux ${\cal F}$ are known 
to order $v^7$ (i.e. 3.5PN order) and given by
\begin{eqnarray}
E & = & -\frac{1}{2}{\eta v^2} \sum_{k=0}^3 E_k v^{2k},\\
{\cal F} & = & \frac{32}{5} \eta^2 v^{10}\sum_{k=0}^7 {\cal F}_k v^k -
\frac{1712}{105} \ln (v) v^{6},
\end{eqnarray}
where, $\eta=m_1m_2/M^2,$ is the symmetric mass ratio
in terms of the total mass $M=m_1+m_2$ 
and  where
the coefficients $E_k$ and ${\cal F}_k$ can be found in 
Ref.~\cite{DIS3}.
Note the presence of the $\log$-term at order $v^6$ in the expression
for the flux.
\begin{figure}[t]
\includegraphics[width=2.5in]{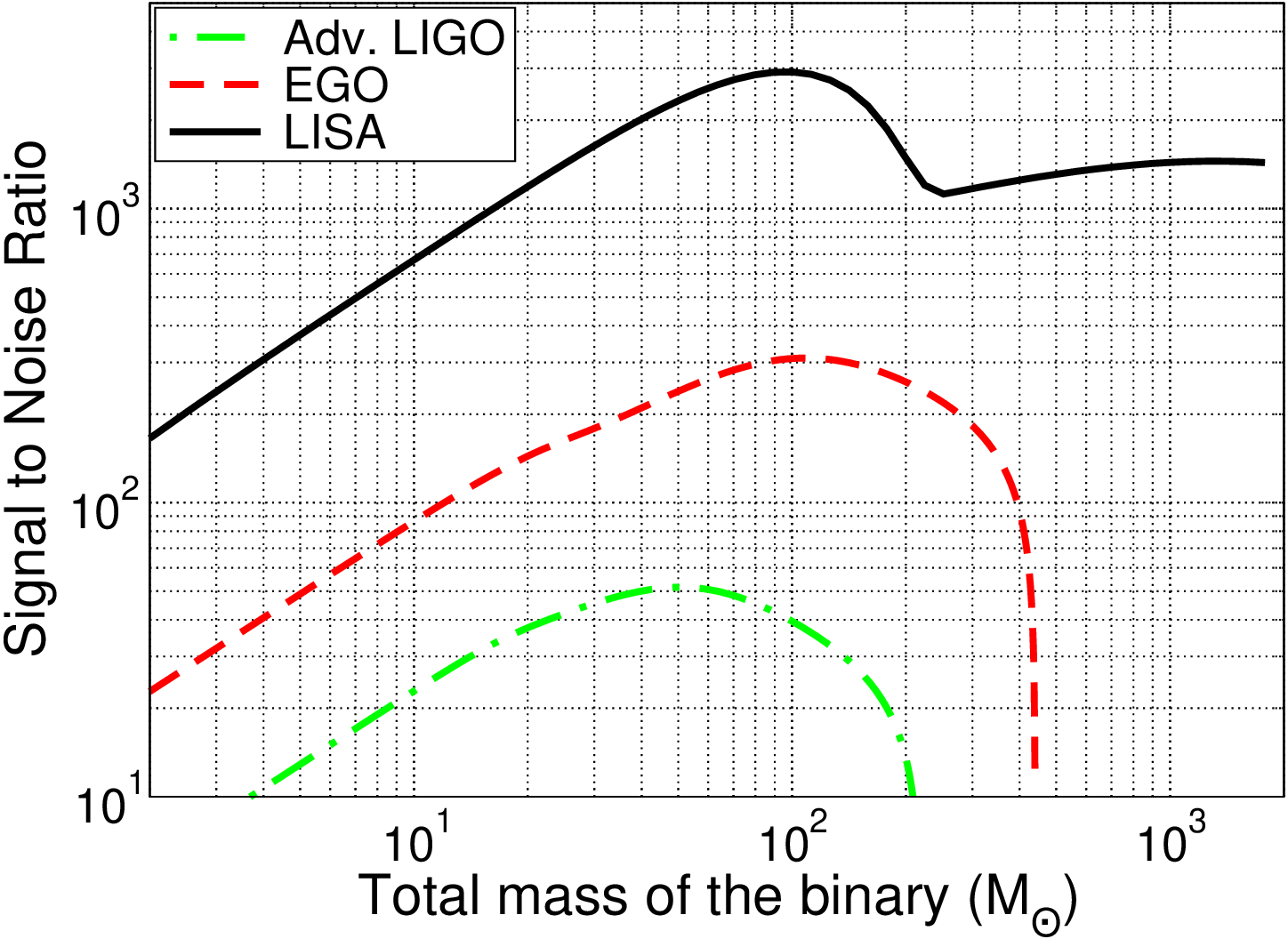}
\caption{The signal-to-noise ratio for stellar mass binary black holes (BBH) in
Advanced LIGO and EGO  and supermassive BBH in
LISA for equal mass binaries at a distance of 200 Mpc
 (for EGO and Advanced LIGO) and $z=1$ (LISA).
In the case of LISA we assume that the signal is integrated for 
a year (last year before coalescence) and in the case of EGO we assume that the signal 
is integrated over a bandwidth from 10 Hz until the binary reaches 
its innermost circular orbit.
The masses of supermassive 
BBH in the case of LISA have been scaled down by a factor of $10^4.$}
\label{fig:snr}
\end{figure}

\begin{figure*}[t]
\includegraphics[width=2.5in]{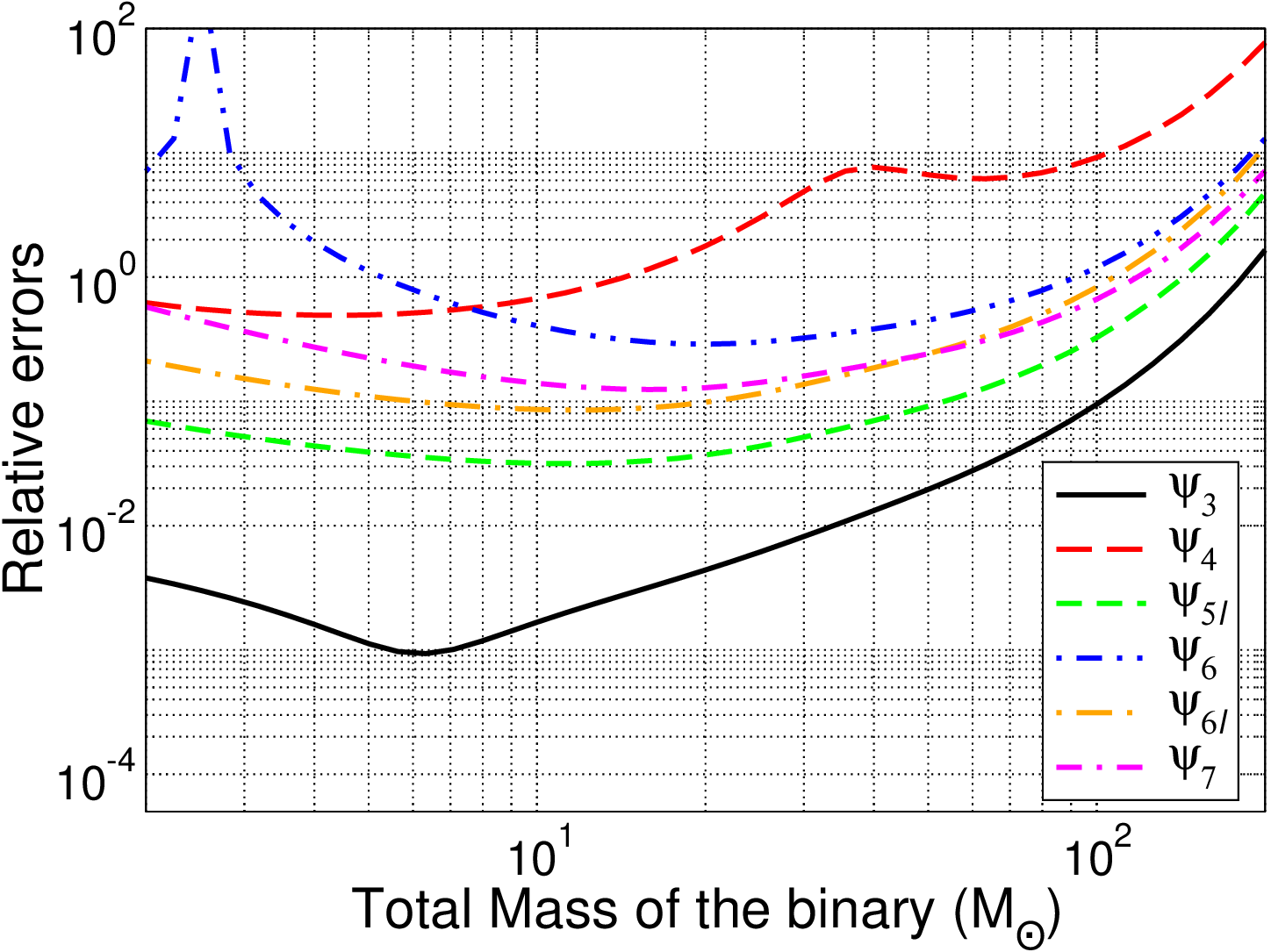}
\hskip 1.0 true cm
\includegraphics[width=2.5in]{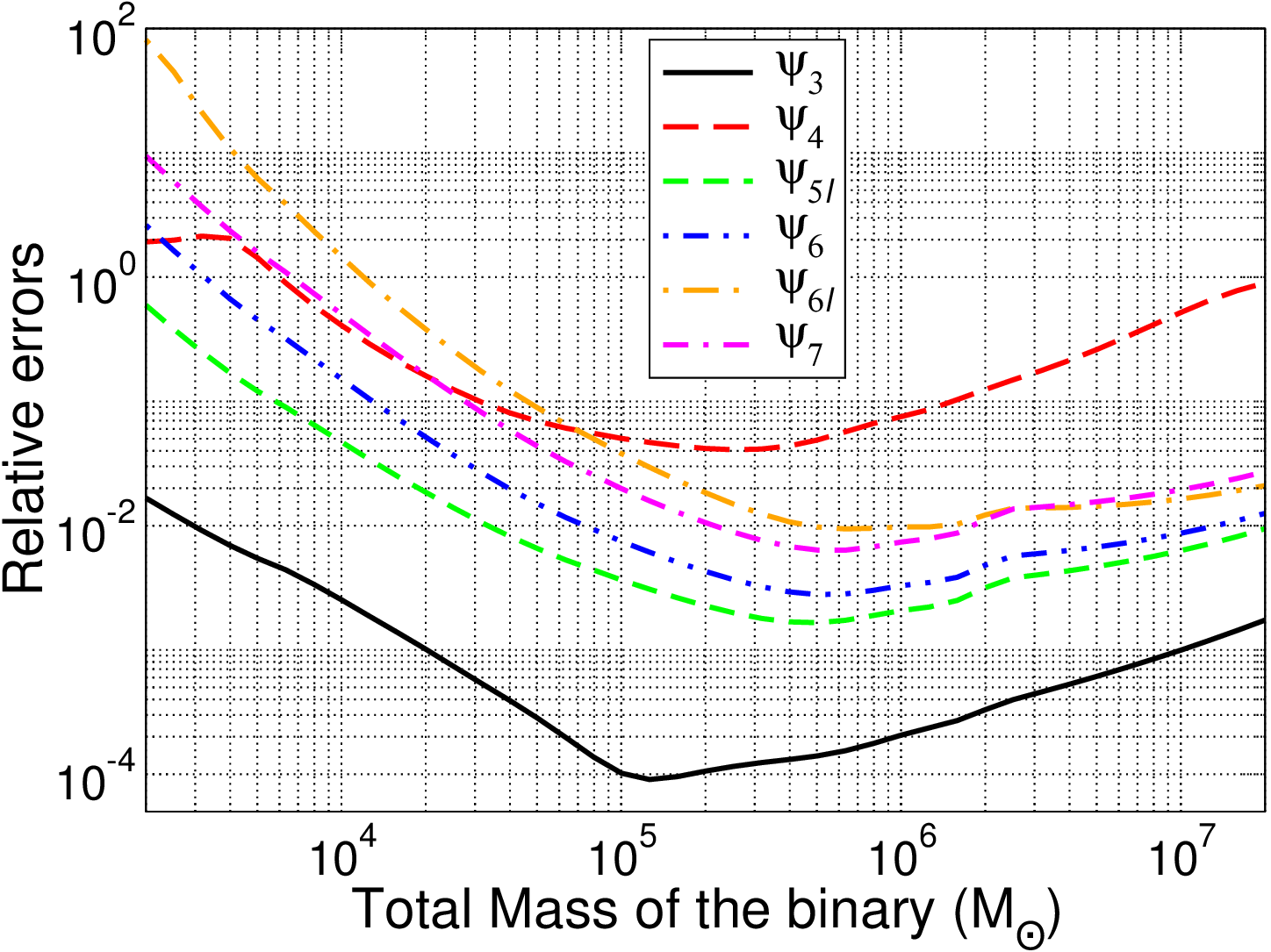}
\caption{Plot showing the relative errors $\Delta \psi_T/\psi_T$, 
in the test parameters $\psi_T=\psi_3,  \psi_4,  
\psi_{5l},  \psi_6,  \psi_{6l},  \psi_7,$ as a function of the total
mass $M$ of a supermassive BBH at a redshift of $z=1$
observed by LISA (right panel) and of a stellar mass compact binary
at a distance of $D_L=200$~Mpc observed by EGO (left panel).  
The rest of the details as in Fig~\ref{fig:snr}.
}
\label{fig:lisa-ego}
\end{figure*}

To understand how we might test the non-linear structure of general
relativity let us begin with the Fourier domain representation $H(f)$ of
the signal from a binary at a luminosity distance $D_L$~\cite{f1}
consisting of black holes of 
masses $m_1$ and $m_2$:
\begin{equation}
H(f) = \frac{C {\cal M}^{5/6}}{D_L\pi^{2/3}} \sqrt{\frac{5}{24}} f^{-7/6}
e^{i\Psi(f) + i \pi/4},
\end{equation}
where ${\cal M} =  \eta^{3/5}M$ is the chirp mass and 
$0\le C \le 1$ is a constant that depends on the relative 
orientation of the detector and source  with a root-mean-square value
of $2/5$ when averaged over all sky locations and source
orientations. 
The phase $\Psi(f)$ is given by
\begin{equation} 
\Psi(f) = 2\pi f t_c - \Phi_c + \sum_{k=0}^7 \left [\psi_k + \psi_{kl}
\ln f \right ] f^{(k-5)/3}.
\label{eq:Fourier Phase}
\end{equation} 
Here $t_c$ and $\Phi_c$ are the epoch of merger and the signal's phase
at that epoch, respectively. 
The non-zero coefficients in the PN expansion of the Fourier 
phase can simply be read off from $\alpha_n$s in Eq.~(3.4) of
 Ref.\ \cite{AISS04}.
The non-vanishing coefficients of the log-terms up to 3.5PN are
$\psi_{5l}$=$-{65\pi\over 384}+{38645\pi\over32256\eta}$
and
$\psi_{6l}$=$-{107\over42\eta}(\pi M)^{1/3}$. For completeness,
$\psi_6$=${3\,\alpha_6\over128\eta}(\pi\,M)^{1/3}-\psi_{{6l}}\log f$.
We have a total of nine post-Newtonian parameters, seven of these
are the coefficients of $v^n$ terms for $n=0,2,3,4,5,6,7,$ and two
are coefficients of $v^n \ln(v)$ terms~\cite{f2} for $n=5,6,$ but each of these
parameters depends only on the masses of the two black holes for the
nonspinning case in GR.

Before proceeding with the  description of our work, let us 
summarise the assumptions implicit in the analysis, the justification
for doing so and their possible implications.
As in most works on this subject,
 to demonstrate  the `principle' of the proposed method 
we  neglect the  effects of spins and eccentricity.
What will change on including  these additional parameters
 is the accuracy of the test. 
The spin effects are relevant only when one of the
black holes is much smaller than the other and/or when the black holes
have their dimensionless spin angular momentum close to unity. It is not clear that astrophysical black holes,
especially the supermassive ones, will be extreme Kerr. Except in cases
where both BHs are extreme Kerr (or close to it) spin effects are less
important for the proposed tests since we have considered black holes
of comparable masses in our study.
The issue of eccentricity, especially for certain LISA sources, is
a complex issue depending on the astrophysical scenario
related to formation mechanisms of the binary. Our
neglect of eccentricity in these cases is a simplifying assumption at present.
Finally let us comment on the use of the so-called restricted PN waveform
in this work.
Not  merely in connection with tests that have been proposed but more
seriously in most  works related to the detection problem in  GW
data analysis,
the late inspiral and merger part is ignored in the first instance. 
One begins by   using
 state-of-the-art restricted PN inspiral templates. 
Restricted PN waveforms will only bring new variety
 (higher harmonics) without
increasing the number of parameters; a full test should definitely use
the full waveform.  Including PN amplitude corrections could
improve the  tests and this is what we are doing as a follow-up
of the present analysis.
By the time LISA and EGO operate there could be reliable merger
waveforms that can be included in the phasing  and this
would  make this test more robust.

Given a high SNR binary black hole event one can, in 
principle, make a model-independent measurement of the above PN coefficients
by accepting those values that best fit the data as our estimates. 
A procedure in which all the parameters 
${\bm \theta} \equiv (t_c,\Phi_c,\psi_k,\psi_{kl}),$
$k=0,2,\ldots,7,$ are independently varied to obtain the best 
possible fit of the signal to the data subjects general relativity to
the most stringent test possible. 
In a recent paper, we explored the power of such a test to determine
all the known coefficients to a relative accuracy of 100\% or
better~\cite{AIQS06a}.
However, this is by no means 
the most powerful test. This is because the covariances between 
the various parameters enhance the errors in their estimation, thereby
diluting the effectiveness of the test.

In the present paper
we have studied the accuracy with which we can measure the 
PN coefficients by treating at a time only three of the nine
$\psi_k$ coefficients to be independent and taking the rest as 
functions of two of the three parameters.
Thus, once a high SNR event is identified, we suggest to fit the 
data to a template wherein  {\it three} terms in the PN
expansion, rather than just {\it two} as in detection problem, (or {\it all} 
the PN terms as proposed in Ref.\ \cite{AIQS06a}), 
are treated as independent parameters.
More precisely, in Einstein's GR, the tests
consist in treating the parameters $\psi_0$ and $\psi_2$ 
as the fundamental ones from which we can measure the masses 
of the two black holes by inverting the relationships $\psi_0=
\psi_0(m_1, m_2)$ and $\psi_2=\psi_2(m_1, m_2),$ and asking if the
measurement of a third parameter, say $\psi_{6l}= \psi_{6l}(m_1,m_2),$
is consistent with the other two.
Instead of the pair $(\psi_0, \psi_2)$ one can, in principle,
equally well take any other pair to be the fundamental set.
The parameters $\psi_0$ and $\psi_2,$ being lower order coefficients,
are best determined as compared to the others and constitute our
favoured pair~\cite{f3}.

We shall consider the estimation of parameters in the ground-based
EGO and space-based LISA, using covariance matrix, for which we assume the
noise PSDs as given in Ref.\ \cite{EGO} and \cite{Barack&Cutler04}, 
respectively. We shall take the fundamental parameters to be
$\psi_0$ and $\psi_2$ in addition to the usual extrinsic parameters
$t_c$  and $\Phi_c$.  We shall take the test
parameter $\psi_T$ to be in turn $\psi_3, \ldots, \psi_7$,
$\psi_{5l}$ and $\psi_{6l}.$ 
It should be noted that there is no test corresponding to the term
involving $\psi_5$ since it has no frequency dependence and simply
redefines of the coalescence phase $\Phi_c$.
For ground-based detectors, Advanced LIGO and EGO, the parameters
include $t_c,\,\Phi_c$ and the three $\psi$'s.
For LISA, on the other hand, the results correspond to the case
of a single detector but with amplitude modulation caused by the
motion of the detector relative to the source. In this case
our Fourier domain waveform will have amplitude, phase and frequency
modulations due to the orbital motion of LISA and we use the waveform
given in Ref~\cite{Cutler98}.
Thus for LISA, in addition to the three $\psi$ parameters related to our
tests we also have the luminosity distance
and the four angles related to  the source's location and orientation.

The power of the tests depends on the SNR achieved
for the source. In Fig.~\ref{fig:snr} we have plotted the SNR in LISA,
EGO and Advanced LIGO~\cite{AISS04}, for BBH binaries at a distance of $z=1$ for LISA 
and a distance of $D_L=200$ Mpc for EGO and Advanced LIGO. 
In the case of EGO, we consider stellar mass
BBH of equal masses with the total mass in the range
$1 M_\odot$ to $400 M_\odot,$ while in the case of LISA the mass range 
is from $10^4 M_\odot$ to $10^7 M_\odot,$ but scaled down by $10^4$ so
as to fit all the curves in the same plot. While the SNR in EGO can reach
several 100's for sources that it might observe every once in a year, in the
case of LISA the SNR could be several 1,000's for the supermassive BBH sources
that it is expected to observe about once per year. The SNR's in both LISA 
and EGO are large enough for the tests to be very powerful probes of
the PN coefficients and the non-linear effects of GR.

The lowest order parameters $\psi_0$ and $\psi_2$ are measured with
the smallest errors. In the case of LISA the errors for a source
at $z=1$ are of order $10^{-5}-10^{-4}$ and in the case of EGO the errors
for a source  at 200 Mpc are of order $10^{-4}-10^{-3}.$
Fig.~\ref{fig:lisa-ego} plots the relative errors $\Delta \psi_T/
\psi_T$ for various parameters $\psi_T$ as a function of the total mass $M.$
From the plots, it is clear that the proposed tests can be performed
 effectively with all $\psi_k$'s, especially in the case of LISA.
This is another reason why LISA is such an important mission.
All the test parameters, including the log-terms at 2.5PN and 3PN order,
can be estimated with fractional accuracies better than $10^{-2}$ 
in the case of LISA for massive BBH binaries with the total mass in the range 
$10^4$-$10^7$ $M_\odot$, and with fractional accuracies better than 100\%
in the case of EGO for stellar mass BBH binaries with the total mass range 
$2$-$10M_\odot$.  This demonstrates the exciting possibility of testing the 
non-linear structure of general relativity using the GW observations by EGO and
LISA.  A similar analysis in the case of Advanced LIGO for sources with the 
total mass $\sim 10M_\odot$, shows that all the parameters, except 
$\psi_4$ and $\psi_{6l},$ can be measured to a relative accuracy of 100\%. 
Thus, though the 3PN log-term cannot be probed with Advanced LIGO, 
the 2.5PN log-term can be tested leading to an interesting possibility 
in the more immediate future.

With reference to Fig.~\ref{fig:lisa-ego},
one may wonder why the error in $\psi_4$ is the largest
 relative to the other, higher order, $\psi$'s.
We believe that there are several reasons for this odd behaviour:
recall that the PN terms in the Fourier phase are given
by $\psi_k f^{(k-5)/3}$.  When $k=5,$ there is no dependence on 
frequency and when $k=4$ the
term varies very slowly as $f^{-1/3}$. Therefore, terms close to $k=5$
are likely to suffer from large variances since the frequency dependence
of the corresponding term is weak.
Although one might expect $\psi_6$ also to suffer from large relative
errors, the fact that in this case the term increases with frequency
as $f^{1/3}$ contributes to making it a more important term than $\psi_4.$ 
We also observe that $\psi_4$ has significantly larger covariances with 
$\psi_0$ and and $\psi_2$ which adds to its poor determination.

\begin{figure}[t]
\includegraphics[width=3.2in]{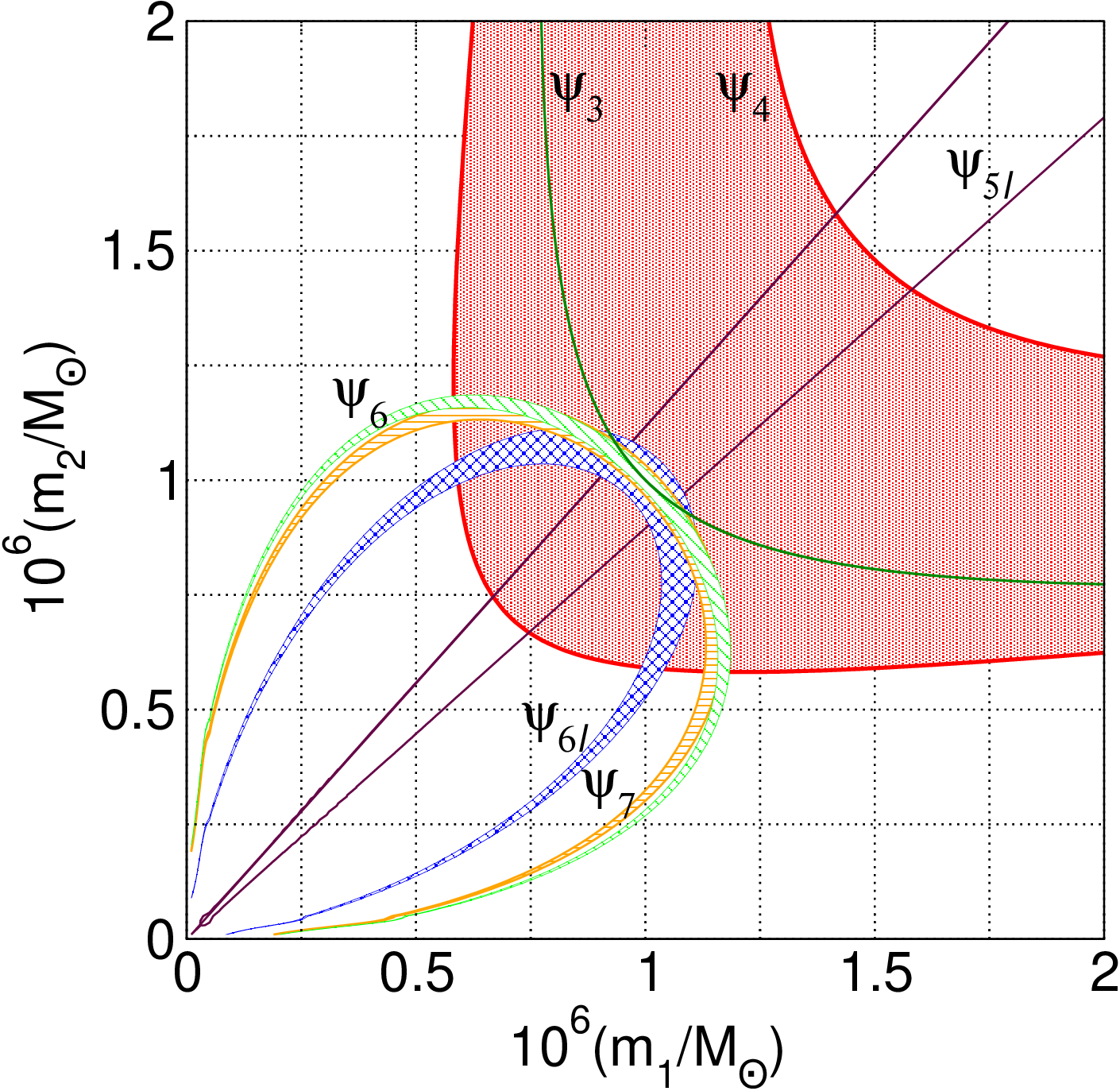}
\caption{Plot showing the regions in the $m_1$-$m_2$ plane that
correspond to 1-$\sigma$ uncertainties 
in the test parameters $\psi_T=\psi_3,  \psi_4,  \psi_{5l},  
\psi_6,  \psi_{6l},  \psi_7$ for a $(10^6,\, 10^6) M_\odot$ supermassive 
black hole binary at a redshift of $z=1$ as observed for a year by LISA.
(Note that the 1-$\sigma$ uncertainty in $\psi_3$ is smaller than 
the thickness of the line.)}
\label{fig:m1m2}
\end{figure}

In Fig~\ref{fig:m1m2}, we have depicted the power of the proposed test
in the $m_1$-$m_2$ plane.
We present the uncertainty contours, with $1$-$\sigma$ error bars, associated with the different test
parameters in the $m_1$-$m_2$ plane, when $\psi_0$ and $\psi_2$ are used
to parametrize the waveform and in the case of LISA.
The parameter $\psi_{6l}$ is much better determined by LISA than EGO, as
one would expect. This figure is an explicit demonstration of the 
efficacy of the proposed test and the accuracy with which the future GW
observations of BH binaries by EGO and LISA
can test GR in its strong field regime.

As mentioned earlier, 
the spin and angular parameters add a lot of structure to the
waveform which contain additional information that can be extracted
and more tests conducted. Covariance between the old and new parameters
is likely to increase the error boxes but the tests become
 more demanding as a result of seeking consistency amongst
 a greater number of parameters.
Future studies should  look into the more general case incorporating the
effects of spin and systematic effects of orbital eccentricity
 that could affect the tests,
 and more interestingly, go beyond the restricted waveform 
approximation by incorporating the amplitude corrections~\cite{ABIQ04} 
to the GW phasing.  

We conclude by discussing the extent to which we can extend the current
proposal to discriminate between different theories of gravity such
as massive graviton theories and scalar-tensor
theories~\cite{Will,DE92}. The limitations of GW phasing to
quantitatively discriminate between alternate theories of gravity has
been critically discussed in~\cite{DE98} and should be kept in mind.
For the massive graviton theories the 1PN phasing term $\psi_2$ is
different and also involves the Compton wavelength of the graviton
$\lambda_g$. Using $\psi_0$ and $\psi_3$ as basic variables and $\psi_2$
as a test~\cite{f4},
we find that bounds can be set on the value of
$\lambda_g$, modulo the neglect of uncomputed higher PN order
corrections in the theory. Using EGO, which will observe stellar mass BH coalescences,
we can set a bound on $\lambda_g$ to be $1.3\times10^{13}$ km whereas with LISA the bounds
are as high as  $7.12\times10^{16}$ km. Scalar-tensor
theories like Brans-Dicke theory, which predicts dipolar
GW emission, have leading additional terms in the the phasing formula at a PN
order lower than in GR. But the dipole GW emission is more important for 
asymmetric binaries than it is for equal mass systems.
However, for such systems spin effects are also expected to play a crucial role.
The present paper deals with only non-spinning binaries and
we postpone the questions relating to dipolar radiation including spin effects
to a future work. Once again, these tests will be limited by the uncomputed higher
order PN contributions in the Brans-Dicke theory.
\acknowledgments
BRI thanks the University of Wales and Cardiff, U.K. for supporting his
visit in January
2006 and  BSS thanks the Raman Research Institute, India for hospitality
during
August 2005 when part of the work was carried out.

{}

\begin{thebibliography}{}
\bibitem{footGc}We use a system of
units in which
$c=G=1.$
\bibitem{f1}
Throughout the paper we assume a cosmological model with zero spatial
curvature
($\Omega_{\kappa}=0$, $\Omega_{\Lambda}+\Omega_{M}=1$) and Hubble's
constant to be $H_0=70$ km s$^{-1}$ Mpc$^{-1}$.
The luminosity distance is given by
$
D_L=\frac{1+z}{H_0}\int_0^z
\frac{dz'}{\left[\Omega_M(1+z')^3+\Omega_\Lambda\right]^{1/2}}\,,
$
where $z$ denotes the redshift of the source.
\bibitem{f2}In the Fourier domain
the $\log$-terms appear at 2.5PN order rather than 3PN because of an
integration that involves $1/f$.
\bibitem{f3}
One could alternatively use the entire PN expansion up to some $k'$
 as a basis, and then check consistency
with $\psi_{k'+1}$ (ignoring higher order terms). This would be relevant
for testing alternative theories where deviations from GR grow larger
with stronger gravity. However, the systematic effects from the neglect
of known higher order PN corrections to GR might contaminate the
difference between the alternate theory and GR in this case.
\bibitem{f4}Given the theoretical status of the phasing
formula in the massive graviton case,
this choice we  implement is the most convenient one  at present.
In a more  complete theory  $\psi_3$
could in general depend on the Compton wavelength as well 
and there would be not much rationale behind the choice proposed. 
One could then  use $\psi_0$ and
$\psi_2$ as basic variables and $\psi_3$ as a test, as before.
\end{thebibliography}
\end{document}